\documentclass[preprint]{elsarticle}

\begin{document}
%\draft
\title{Intensive variables in the framework of the non-extensive thermostatistics}
\author{A.M. Scarfone}
\ead{antonio.scarfone@polito.it}
\address{Istituto di Fisica della Materia (INFM--CNR) c/o Dipartimento di Fisica, Politecnico di
Torino\\ Corso Duca degli Abruzzi 24, I-10129, Torino, Italy.}

\date{\today}
\begin{abstract}
By assuming an appropriate energy composition law between two systems governed by the same non-extensive entropy, we revisit the definitions of temperature and pressure, arising from the zeroth principle of thermodynamics, in a manner consistent with the thermostatistics structure of the theory. We show that the definitions of these quantities are sensitive to the composition law of entropy and internal energy governing the system. In this way, we can clarify some questions raised about the possible introduction of intensive variables in the context of non-extensive statistical mechanics.
\end{abstract}

\maketitle

%%%%%%%%%%%%%%%%%%%%%%%%%%%%%%%%%%%%%%%%%%%%%%%%%%%%%%%%%%%%%%%%%%%%%%%%%%%%%%%%%%%%%%%%%%%%%%%%%%%%%%%%%%%%%%%%%%%%%%%
\section{Introduction}

The $q$-entropy, in the $(2-q)$--re-parameterized form \cite{Scarfone1,Wada1,Scarfone5,Scarfone6} (hereinafter denoted by $S_{2-q}$), reads
\begin{eqnarray}
S_{2-q}=\sum_{i=1}^W{p_i^{2-q}-p_i\over q-1} \ ,\label{Tsallis}
\end{eqnarray}
where $p\equiv\{p_i\}_{i=1,\ldots,W}$ is a discrete probability distribution function for the $W$ distinguishable energy levels and $q$ is a real deformation parameter whose value should be such to guarantee the existence of all the relevant expectation values of the theory.\\ Accounting for the normalization of the distribution, the entropic form (\ref{Tsallis}), with the index $q$ re-parameterized in $2-q$, becomes
\begin{eqnarray}
S_q={\sum_{i=1}^Wp_i^q-1\over 1-q} \ ,
\end{eqnarray}
that is the Tsallis entropy proposed in 1988 \cite{Tsallis}, although the original version of Tsallis entropy was introduced previously in information theory by Dar\'{o}czy \cite{Daroczy} in the form $S^{\rm D}_\alpha={\sum_ip_i^\alpha-1\over2^{1-\alpha}-1}$ and appears in a quantum version in the Wehrl review  \cite{Wehrl} with the form $S_\alpha^{\rm W}={{\rm Tr}\rho^\alpha-1\over1-\alpha}$.\\
The re-parameterized form (\ref{Tsallis}) has been obtained, in a unified fashion, as a particular case of the two-parameters family of Sharma-Taneja-Mittal entropies \cite{Scarfone1} and
it has the advantage to present the so-called first formulation of the non-extensive statistical mechanics \cite{Tsallis1} in a little bit easily way. Beyond this aesthetics question, the formalism adopted in this work is fully equivalent with the first formulation quoted above.\\
Entropy $S_{2-q}$ represents a possible alternative to the Boltzmann-Gibbs entropy (recovered in the $q\to1$ limit: $S_1=-\sum_ip_i\,\ln p_i$), in the study of anomalous statistical systems, especially when they are plagued by long range interactions or memory effects persistent in time (see \cite{Tsallis0} and references therein).\\
One of the most stimulating proprieties of the $q$-entropy is its composition law that, for statistically independents systems, assumes the form
\begin{eqnarray}
S_{2-q}^{{\rm A}\cup{\rm B}}=S_{2-q}^{\rm A}+S_{2-q}^{\rm B}+(q-1)\,S_{2-q}^{\rm A}\,S_{2-q}^{\rm B} \ . \label{comp}
\end{eqnarray}
This property has been employed in the definition of intensive quantities like temperature \cite{Abe,Abe1,Abe2,Wada,Wada3,Wada4,Scarfone0}, considering
a typical situation in which the system, at thermal equilibrium
with energy $U^{{\rm A}\cup{\rm B}}=U^{\rm A}+U^{\rm B}$, is ideally separated in two statistically independent parts. In this way, one is neglecting the interactions among the components of the system with respect to the bulk energy, although a kind of interaction between the parts of the system is required to allow thermal equilibrium.\\
By supposing that the equilibrium state is the one that maximizes the entropy of the system, under the condition of constant internal energy $dU^{\rm A}=-dU^{\rm B}$, one can introduce the temperature, according to the zeroth principle of thermodynamics, as
\begin{eqnarray}
\bar T\propto\Big[1+(q-1)\,S_{2-q}\Big]\,\left({\partial U\over\partial S_{2-q}}\right)_V \ ,\label{temp00}
\end{eqnarray}
where $V$ is the volume of the system, that coincides with the separation constant between the two sub-systems.\\
This method has been criticized by some authors \cite{Wang,Nauenberg}, in particular for what concerns the hypothesis of statistical independence and the resulting Eq. (\ref{comp}), that, as known, is incompatible with the assumption of additivity of energy. More exactly, supposing that $U^{{\rm A}\cup{\rm B}}=U^{\rm A}+U^{\rm B}$, the equilibrium distribution, when the system is governed by the entropy $S_{2-q}$, factorizes in
\begin{eqnarray}
p^{{\rm A}\cup{\rm B}}=p^{\rm A}\,p^{\rm B}\,(1+\delta p) \ ,
\end{eqnarray}
where $\delta p$ takes into account the statistical correlations among the two sub-systems.\\
Therefore, the assumption of statistical independence corresponds to ignore the correlations that are at the origin of the non-extensive proprieties of the system. This conclusion is also in agreement with the results reported in \cite{Toral} where it has been shown that, in the microcanonical picture, Eq. (\ref{temp00}) coincides with the well-known Boltzmann definition of temperature which is consistent with a multiplicative structure for the number of configurations $\Omega({\rm A}+{\rm B})=\Omega({\rm A})\,\Omega({\rm B})$ that, again, corresponds to ignore the possible correlations/interactions among the two sub-systems.\\
In this letter, we show that, assuming an appropriate interaction between the parts of the system, we can avoid the above question and we are able to introduce intensive variables, like temperature and pressure, in a manner consistent with the mathematical structure of the theory.\\
The letter is organized as following. In the next section 2, we revisit the equilibrium distribution that maximizes the entropy $S_{2-q}$. In the section 3, we introduce the intensive variables, by means of the zeroth law of thermodynamics, in a way compatible with the factorization of the distribution. This fact will be shown in the successive section 4. The conclusion is reported in the final section 5.

%%%%%%%%%%%%%%%%%%%%%%%%%%%%%%%%%%%%%%%%%%%%%%%%%%%%%%%%%%%%%%%%%%%%%%%%%%%%%%%%%%%%%%%%%%%%%%%%%%%%%%%%%%%%%%%%%%%%%%%
\section{Equilibrium distribution}

According to the maximal entropy principle, the equilibrium distribution coincides with the one that maximizes the entropy under appropriate constraints representing different macroscopic observable. In the canonical picture \cite{Pathria}, these constraints are given by the mean energy $U=\sum_i p_i\,\epsilon_i$ and the normalization $\sum_i p_i=1$. In this way, from entropy (\ref{Tsallis}) one obtains the following distribution
\begin{eqnarray}
p_i={1\over \tilde Z_{2-q}}\,\exp_q\left(-\tilde\beta_{2-q}\,\epsilon_i\right)\equiv{1\over\bar Z_{2-q}}\,\exp_q\left(-\bar\beta_{2-q}\left(\epsilon_i-U\right)\right) \ .\label{pdf}
\end{eqnarray}
The $q$-exponential, $\exp_q(x)=1/\exp_{2-q}(-x)$, defined by
\begin{eqnarray}
\exp_q(x)=\left[1+(1-q)\,x\right]_+^{1\over1-q} \ ,
\end{eqnarray}
with $[x]_+=x$ for $x\geq0$ and $[x]_+=0$ for $x\leq0$, is the inverse function of the $q$-logarithm, $\ln_q(x)=-\ln_{2-q}(1/x)$, given by
\begin{eqnarray}
\ln_q(x)={x^{1-q}-1\over1-q} \ .
\end{eqnarray}
Both these functions appear recurrently in the thermostatistics theory based on the entropy $S_{2-q}$.\\
Distribution (\ref{pdf}) differs from the Gibbs one for its asymptotic power law behavior, whereas the exponential tail, typical of the Gibbs distribution, is obtained in the $q\to1$ limit, in agreement with the relation $\exp_1(x)=\exp(x)$.\\
As known, distribution (\ref{pdf}) for $q<1$ shows a cut-off in the energy spectrum given by the condition $\epsilon_i\leq\epsilon_{\rm cutoff}=1/[(1-q)\,\tilde\beta_{2-q}]$. This feature of the distribution introduces interesting prospective about the potential applications of the theory although it is sometime debated in literature \cite{Plastino,Bagci1,Bagci2}.

Quantities $\tilde Z_{2-q}$, $\tilde\beta_{2-q}$ and $\bar Z_{2-q}$, $\bar\beta_{2-q}$ are given by
\begin{eqnarray}
\beta=(2-q)\,\tilde\beta_{2-q}\,\left(\tilde Z_{2-q}\right)^{q-1}=(2-q)\,\bar\beta_{2-q}\,\left(\bar Z_{2-q}\right)^{q-1} \ ,\label{b1}
\end{eqnarray}
and
\begin{eqnarray}
\gamma=\ln_{2-q}(\alpha\,\tilde Z_{2-q})=\ln_{2-q}(\alpha\,\bar Z_{2-q})-\beta\,U \ ,\label{g1}
\end{eqnarray}
where $\beta$ and $\gamma$ are the Lagrange multipliers associated with the mean energy and the normalization, respectively. They can be determined by the corresponding constraint equations.\\
In Eq. (\ref{g1}), we posed $\alpha=(2-q)^{1/(q-1)}$, a constant that is reducing to $e^{-1}$ in the $q\to1$ limit.\\ We can also verify the following relations
\begin{eqnarray}
\bar Z_{2-q}=\exp_{2-q}\left(S_{2-q}\right) \ ,\label{barz}
\end{eqnarray}
and
\begin{eqnarray}
\tilde\beta_{2-q}={\bar\beta_{2-q}\over1+(1-q)\,\bar\beta_{2-q}\,U} \ ,\label{b2}
\end{eqnarray}
that will be used later in this letter.\\
Although the two different expressions in Eq. (\ref{pdf}) are equivalent each to the other, the second has been mostly used in the development of the theory, mainly because of its explicit invariance for the origin energy translation \cite{Tsallis1}; a fact which, although valuable in terms of the elegance of the theory, has no practical implications if one considers that the origin of energy can be ultimately fixed according to the third principle of thermodynamics.\\ Notwithstanding, in spite of its form, it can be shown that also the first expression in Eq. (\ref{pdf}) is invariant under uniform translation of the energy (see appendix A for a proof), a fact often overlooked in literature. As will be shown in the following, the first expression in Eq. (\ref{pdf}) can be used fruitfully to derive, in a concise way, some proprieties of the theory under inspection.\\
In conclusion of this section, let us recall the fundamental relation
\begin{eqnarray}
\left({\partial S_{2-q}\over\partial U}\right)_V=\beta \ ,\label{Leg1}
\end{eqnarray}
that plays a role in the introduction of the Legendre structure of the theory \cite{Wada1}. Equation (\ref{Leg1}), can be obtained from the relation $\sum_i dp_i=0$, that follows from the normalization of the distribution, and from the first law of thermodynamics $dU=\sum_i\epsilon_i\,dp_i$ with the no-work condition ($\delta{\cal L}=\sum_i p_i\,d\epsilon_i=0$), valid, for instance, in transformations at constant volume. In fact, by differentiating the entropy $S_{2-q}$, we obtain
\begin{eqnarray}
\nonumber
dS_{2-q}&=&{2-q\over q-1}\sum_{i=1}^Wp_i^{1-q}\,dp_i={2-q\over q-1}\sum_{i=1}^W\left(\tilde Z_{2-q}\right)^{q-1}\,\left[1-(1-q)\,\tilde\beta_{2-q}\,\epsilon_i\right]\,dp_i\\
&=&(2-q)\,\left(\tilde Z_{2-q}\right)^{q-1}\,\tilde\beta_{2-q}\sum_{i=1}^W\epsilon_i\,dp_i=\beta\,dU \ ,
\end{eqnarray}
where we used Eq. (\ref{b1}).

%%%%%%%%%%%%%%%%%%%%%%%%%%%%%%%%%%%%%%%%%%%%%%%%%%%%%%%%%%%%%%%%%%%%%%%%%%%%%%%%%%%%%%%%%%%%%%%%%%%%%%%%%%%%%%%%%%%%%%%
\section{Intensive variables}

According to the zeroth principle of thermodynamics, a possible definition of temperature can be achieved through the following three assumptions: a) the equilibrium distribution $p_{\rm eq}\equiv p^{{\rm A}\cup{\rm B}}$ is the one that maximizes the entropy of the system, that means $dS=0$; b) supposing to divide ideally the system into two arbitrary parts A and B, we assume that it is known a composition rule for the entropy, i.e. $S^{{\rm A}\cup{\rm B}}\equiv S[p^{{\rm A}\cup{\rm B}}]=f(S^{\rm A},\,S^{\rm B})$; c) assuming the system isolated from the environment and supposing to know the composition law of energy, according to $U^{{\rm A}\cup{\rm B}}=g(U^{\rm A},\,U^{\rm B})$, we have $dU^{{\rm A}\cup{\rm B}}=dg(U^{\rm A},\,U^{\rm B})=0$. In this way, one can identify the temperature with the separation constant (modulo an arbitrary multiplicative constant, fixing the thermometric scale) between the two parts in which the system has been divided.

In the ordinary thermostatistics, the points b) and c) are realized by posing $f(x,\,y)\equiv g(x,\,y)=x+y$. As known, these assumptions are consistent with the exponential form of the Gibbs distribution, because
\begin{eqnarray}
\nonumber
p_{ij}^{{\rm A}\cup{\rm B}}&\propto&\exp\left(-\beta\,g\left(\epsilon_i^{\rm A},\,\epsilon_j^{\rm B}\right)\right)\\
&=&\exp\left(-\beta\,\epsilon_i^{\rm A}\right)\,\exp\left(-\beta\,\epsilon_j^{\rm B}\right)\propto p_i^{\rm A}\,p_j^{\rm B} \ ,
\end{eqnarray}
and
\begin{eqnarray}
S^{{\rm A}\cup{\rm B}}=f\left(S^{\rm A},\,S^{\rm B}\right)=S^{\rm A}+S^{\rm B} \ .
\end{eqnarray}
In this way, according to a), we have
\begin{eqnarray}
dS^{{\rm A}\cup{\rm B}}=\left({\partial S^{\rm A}\over\partial U^{\rm A}}\right)_VdU^{\rm A}+\left({\partial S^{\rm B}\over\partial U^{\rm B}}\right)_VdU^{\rm B}=0 \ ,
\end{eqnarray}
so that, from the condition $dU^{\rm A}=-dU^{\rm B}$, one introduces the temperature as
\begin{eqnarray}
{1\over T}\propto\left({\partial S\over\partial U}\right)_V\equiv\beta \ .\label{temp0}
\end{eqnarray}

Similarly, in the framework of the non-extensive thermostatistics, assumptions b) and c) are realized by posing $f(x,\,y)=x+y+(q-1)\,x\,y$ and $g(x,\,y)=x+y$, respectively. In this way, we obtain
\begin{eqnarray}
\nonumber
dS_{2-q}^{{\rm A}\cup{\rm B}}&=&\left[1+(q-1)\,S_{2-q}^{\rm B}\right]\,\left({\partial S_{2-q}^{\rm A}\over\partial U^{\rm A}}\right)_VdU^{\rm A}\\
&+&\left[1+(q-1)\,S_{2-q}^{\rm A}\right]\,\left({\partial S_{2-q}^{\rm B}\over\partial U^{\rm B}}\right)_VdU^{\rm B}=0 \ ,\label{comp1}
\end{eqnarray}
and the definition of the temperature becomes
\begin{eqnarray}
{1\over \bar T}\propto{\beta\over1+(q-1)\,S_{2-q}} \ ,
\end{eqnarray}
that coincides, modulo the pre-factor $1/(2-q)$, with the quantity $\bar\beta_{2-q}$ appearing in the second expression of Eq. (\ref{pdf}).\\
As already discussed in the section 1, assumption b) is consistent, in this case, only if the equilibrium distribution factorizes in $p_{ij}^{{\rm A}\cup{\rm B}}=p_i^{\rm A}\,p_j^{\rm B}$, which is certainly not true if the internal energies are linearly additive, due to the non-exponential form of the distribution.

A way to overcome this difficulty is to modify the assumption c), by postulating an interaction among the parts of the system. Hereinafter, we assume for the internal energy the following expression
\begin{eqnarray}
U^{{\rm A}\cup{\rm B}}=U^{\rm A}+U^{\rm B}+\omega\,U^{\rm A}\,U^{\rm B} \ ,\label{energy}
\end{eqnarray}
where $\omega$ is a constant. This relation is clearly associative, that means
\begin{eqnarray}
U^{{\rm A}\cup{\rm B}\cup{\rm C}}=U^{({\rm A}\cup{\rm B})\cup{\rm C}}=U^{{\rm A}\cup({\rm B}\cup{\rm C})} \ .
\end{eqnarray}
Without lost of generality, we pose
\begin{eqnarray}
(q-1)\,\tilde\beta_{2-q}=\omega \ ,\label{parr}
\end{eqnarray}
so that, by using relations (\ref{b1}) and (\ref{g1}), we can write
\begin{eqnarray}
q=1+{\omega\over\beta-\omega\,\gamma} \ ,
\end{eqnarray}
that establishes implicitly a link between the deformed parameter and the coupling constant of the interaction, a fact already observed in others situations \cite{Du1,Du2,Du3,Scarfone3,Scarfone4,Scarfone2}.
From Eq. (\ref{parr}) it follows also that systems governed by an attrattive interaction ($\omega<0$) are described by a $q<1$ otherwise systems governed by a repulsive interaction ($\omega>0$) are described by a $q>1$. Therefore, the emerging cut-off in the distribution for $q<1$ is consistent with the confining nature of the attrattive interactions.

As will be shown in the next section, Eq. (\ref{energy}) with the position (\ref{parr}), is consistent with the hypothesis of factorization of the distribution (\ref{pdf}). For now, we just discuss the consequences of Eq. (\ref{energy}) on the definition of intensive variables such as temperature and pressure.\\
In fact, since the internal energy is constant, we can write
\begin{eqnarray}
{dU^{\rm A}\over1+\omega\,U^{\rm A}}=-{dU^{\rm B}\over1+\omega\,U^{\rm B}} \ ,
\end{eqnarray}
that plugged into Eq. (\ref{comp1}), gives us the following definition of the temperature
\begin{eqnarray}
{1\over\tilde T}\propto{1+\omega\,U\over1+(q-1)\,S_{2-q}}\,\beta \ .\label{temp}
\end{eqnarray}
With the help Eq. (\ref{b2}) we can transform the right hand side of this equation in the quantity
$\tilde\beta_{2-q}$ (modulo the pre-factor $1/(2-q)$) that appears in the first expression of the distribution (\ref{pdf}). Thus, we pose consistently
\begin{eqnarray}
{1\over\tilde T}=\tilde\beta_{2-q} \ ,\label{temp1}
\end{eqnarray}
that recovers, in the $q\to1$ limit, the definition of temperature of the Boltzmann-Gibbs theory.

Similar arguments can be applied to introduce other physical intensive variables like the pressure. In fact, assuming  a dependence of the energy levels from the volume of the system, we obtain
\begin{eqnarray}
\tilde P\propto{\tilde T\over1+(q-1)\,S_{2-q}}\,\left({\partial\,S_{2-q}\over\partial V}\right)_U\equiv-{1\over1+\omega\,U}\,\left({\partial\,U\over\partial V}\right)_S \ ,\label{pres}
\end{eqnarray}
where we used the relation
\begin{eqnarray}
\left({\partial U\over\partial S_{2-q}}\right)_V\,\left({\partial S_{2-q}\over\partial V}\right)_U=-\left({\partial U\over\partial V}\right)_S \ .
\end{eqnarray}
It is clear that the interaction modifies the definition of pressure that, otherwise, would maintain the same form of the standard theory $P=-(\partial U/\partial V)_S$, even when the entropy is expressed in terms of $S_{2-q}$ \cite{Scarfone0}.

%%%%%%%%%%%%%%%%%%%%%%%%%%%%%%%%%%%%%%%%%%%%%%%%%%%%%%%%%%%%%%%%%%%%%%%%%%%%%%%%%%%%%%%%%%%%%%%%%%%%%%%%%%%%%%%%%%%%%%%
\section{Consistence with the thermostatistics theory}

In order to show the consistence of the composition law (\ref{energy}) with the factorization of the distribution, we begin by considering the following two-body energy spectrum
\begin{eqnarray}
\epsilon_{ij}^{{\rm A}\cup{\rm B}}=\epsilon_i^{\rm A}+\epsilon_j^{\rm B}+\omega\,\epsilon_i^{\rm A}\epsilon_j^{\rm B} \ ,\label{sume}
\end{eqnarray}
that, clearly, reproduces Eq. (\ref{energy}) when $p^{{\rm A}\cup{\rm B}}$ factorizes in the product of two statistically independent distributions $p^{\rm A}$ and $p^{\rm B}$.\\
It is worthy to observe that a similar interaction term has been advanced also in others situations like, for instance, in \cite{Algin1} where, in the framework of the quantum algebra formalism, Authors derived several thermodynamic functions of a Fermion gas whose Hamiltonian is reminiscent of what is reported in Eq. (\ref{sume}). Furthermore, possible implications of a such Hamiltonian with the non-extensive thermostatistics and the Fibonacci calculus has been advanced in \cite{Algin2}.\\
According to Eq. (\ref{sume}) it is immediate to verify that
\begin{eqnarray}
\nonumber
\left[1-(1-q)\,\tilde\beta_{2-q}\,\epsilon_{ij}^{{\rm A}\cup{\rm B}}\right]^{1\over1-q}
&=&\left[1-(1-q)\,\tilde\beta_{2-q}\,\left(\epsilon_i^{\rm A}+\epsilon_j^{\rm B}+\omega\,\epsilon_i^{\rm A}\epsilon_j^{\rm B}\right)\right]^{1\over1-q}\\
\nonumber
&=&\left[1-(1-q)\,\tilde\beta_{2-q}\,\epsilon_i^{\rm A}\right]^{1\over1-q}\\
&\times&\left[1-(1-q)\,\tilde\beta_{2-q}\,\epsilon_j^{\rm B}\right]^{1\over1-q} \ , \label{facte}
\end{eqnarray}
when Eq. (\ref{parr}) holds.
On the other hand, by recalling that
\begin{eqnarray}
\tilde Z_{2-q}=\sum_{i=1}^W\left[1-(1-q)\,\tilde\beta_{2-q}\,\epsilon_i\right]^{1\over1-q} \ ,
\end{eqnarray}
we obtain the further relation
\begin{eqnarray}
\tilde Z_{2-q}^{{\rm A}\cup{\rm B}}=\tilde Z_{2-q}^{\rm A}\,\tilde Z_{2-q}^{\rm B} \ ,
\end{eqnarray}
which, together with the formula (\ref{facte}), gives
\begin{eqnarray}
p_{ij}^{{\rm A}\cup{\rm B}}=p_i^{\rm A}\,p_j^{\rm B} \ .
\end{eqnarray}
We observe that Eq. (\ref{sume}) is consistent with the cut-off condition for $q<1$ since it surely holds for the whole system whenever is satisfied for each single part. In fact, by using Eq. (\ref{parr}), the cut-off condition becomes
\begin{eqnarray}
\epsilon_i^{\rm X}\leq\epsilon_{\rm cutoff}^{\rm X}=1/|\omega| \ ,\label{cc}
\end{eqnarray}
where X stands for A or B, so that
\begin{eqnarray}
\nonumber
1-|\omega|\,\epsilon_{ij}^{{\rm A}\cup{\rm B}}&=&\left(1-|\omega|\,\epsilon_i^{\rm A}\right)\,\left(1-|\omega|\,\epsilon_j^{\rm B}\right)\\
\nonumber
&\geq&\left(1-|\omega|\,\epsilon_{\rm cutoff}^{\rm A}\right)\,\left(1-|\omega|\,\epsilon_{\rm cutoff}^{\rm B}\right)\\
&\geq&0 \ ,
\end{eqnarray}
which implies
\begin{eqnarray}
\epsilon_{ij}^{{\rm A}\cup{\rm B}}<\epsilon_{\rm cutoff}^{{\rm A}\cup{\rm B}} \ ,
\end{eqnarray}
with
\begin{eqnarray}
\epsilon_{\rm cutoff}^{{\rm A}\cup{\rm B}}=\epsilon_{\rm cutoff}^{\rm A}+\epsilon_{\rm cutoff}^{\rm B}-|\omega|\,\epsilon_{\rm cutoff}^{\rm A}\,\epsilon_{\rm cutoff}^{\rm B} \ .
\end{eqnarray}
Furthermore, from condition (\ref{cc}) it follows easily that
\begin{eqnarray}
1+\omega\,U>0 \ ,
\end{eqnarray}
both for the attractive case that for the repulsive case, since the mean value of a quantity is surely less or equal than its maximum value. This fact assures the consistence of the definitions (\ref{temp}) and (\ref{pres}).

Another interesting relation can be derived considering, for instance, the variation of $U^{{\rm A}\cup{\rm B}}$ at constant $U^{\rm B}$ (risp. constant $U^{\rm A}$). In this case, recalling Eq. (\ref{Leg1}), we obtain
\begin{eqnarray}
\beta^{{\rm A}\cup{\rm B}}=\beta^{\rm A}\,{1+(q-1)\,S_{2-q}^{\rm B}\over1+\omega\,U^{\rm B}} \ ,
\end{eqnarray}
and, taking into account Eqs. (\ref{barz}) and (\ref{b2}), it can be rewritten as
\begin{eqnarray}
\tilde\beta_{2-q}^{{\rm A}\cup{\rm B}}=\tilde\beta_{2-q}^{\rm A} \ ,
\end{eqnarray}
in accordance with the intensive nature of $\tilde\beta_{2-q}$.

It should be noted that, based on the equivalence of the two expressions in Eq. (\ref{pdf}), the factorization of the first form implies the factorization of the second. In other words, Eq. (\ref{sume}) also assures
\begin{eqnarray}
\nonumber
& &{1\over\bar Z_{2-q}^{{\rm A}\cup{\rm B}}}\left[1-(1-q)\,\bar\beta^{{\rm A}\cup{\rm B}}_{2-q}\,\left(\epsilon_{ij}^{{\rm A}\cup{\rm B}}-U^{{\rm A}\cup{\rm B}}\right)\right]^{1\over1-q}\\
\nonumber
&=&{1\over\bar Z_{2-q}^{\rm A}}\left[1-(1-q)\,\bar\beta^{\rm A}_{2-q}\left(\epsilon_i^{\rm A}-U^{\rm A}\right)\right]^{1\over1-q}\\
&\times&{1\over\bar Z_{2-q}^{\rm B}}\left[1-(1-q)\,\bar\beta^{\rm B}_{2-q}\left(\epsilon_j^{\rm B}-U^{\rm B}\right)\right]^{1\over1-q} \ ,
\end{eqnarray}
although, in this case, the distributions have all different $\bar\beta_{2-q}$, related by
\begin{eqnarray}
\bar\beta^{{\rm A}\cup{\rm B}}_{2-q}={\bar\beta^{\rm A}}_{2-q}\,\left[1+(1-q)\,\bar\beta^{\rm B}_{2-q}\,U^{\rm B}\right]={\bar\beta^{\rm B}}_{2-q}\,\left[1+(1-q)\,\bar\beta^{\rm A}_{2-q}\,U^{\rm A}\right] \ ,
\end{eqnarray}
since these are not intensive quantities.

We can explore further the intensive nature of temperature and pressure in other contexts. For examples, we can consider a system composed by $N$ identical monades, all at the same mean energy $U^{\cal M}$, described by the same entropy $S_{2-q}^{\cal M}$ and interacting according to Eq. (\ref{sume}). This means that the energy spectrum of the whole system will be given by
\begin{eqnarray}
\epsilon_{i_1,i_2,\ldots i_N}=\sum_{j=1}^N\epsilon_{i_j}+\omega\sum_{j,k=1}^N\epsilon_{i_j}\,\epsilon_{i_k}+\omega^2
\sum_{j,k,l=1}^N\epsilon_{i_j}\,\epsilon_{i_k}\,\epsilon_{i_l}+\ldots \ .
\end{eqnarray}
Clearly, for $\omega\ll1$, the only significative contribute cames from the two-body interaction, although, for sake of rigor, we must consider the whole expression.\\
In this way, the equilibrium distribution of the system is
\begin{eqnarray}
p_{i_1,i_2,\ldots i_N}={1\over\tilde Z_{2-q}^{N\cal M}}\,\exp_q\left(-\tilde\beta_{2-q}^{N\cal M}\,\epsilon_{i_1,i_2,\ldots i_N}\right) \ ,
\end{eqnarray}
where, we assume $\omega=(1-q)\,\tilde\beta_{2-q}^{N\cal M}$. This formula factorizes in
\begin{eqnarray}
p_{i_1,i_2,\ldots i_N}=\prod_{j=1}^Np_{i_j} \ ,
\end{eqnarray}
being
\begin{eqnarray}
p_{i_j}={1\over\tilde Z_{2-q}^{\cal M}}\,\exp_q\left(-\tilde\beta_{2-q}^{N\cal M}\,\epsilon_{i_j}\right) \ ,
\end{eqnarray}
which implies the following relation
\begin{eqnarray}
\tilde Z_{2-q}^{N{\cal M}}=\left(\tilde Z^{\cal M}_{2-q}\right)^N \ ,
\end{eqnarray}
as it can be easily verified.
Furthermore, we have
\begin{eqnarray}
U^{N{\cal M}}={1\over\omega}\Big[\left(1+\omega\,U^{\cal M}\right)^N-1\Big] \ ,\label{me}
\end{eqnarray}
and
\begin{eqnarray}
S_{2-q}^{N{\cal M}}={1\over q-1}\Big\{\left[1+(q-1)\,S_{2-q}^{\cal M}\right]^N-1\Big\} \ ,\label{se}
\end{eqnarray}
that express the mean value $U^{N{\cal M}}$ and the entropy $S_{2-q}^{N{\cal M}}$ of the whole system in terms of the corresponding quantities of the individual monads.\\
These two equations can be used, recalling Eq. (\ref{Leg1}), to obtain the relation
\begin{eqnarray}
\beta^{N{\cal M}}=\beta^{\cal M}\left(1+\omega\,U^{\cal M}\right)^{1-N} \ ,\label{sb}
\end{eqnarray}
that, taking into account the definition (\ref{b1}) and Eq. (\ref{b2}), gives
\begin{eqnarray}
\tilde\beta^{N{\cal M}}_{2-q}=\tilde\beta_{2-q}^{\cal M} \ ,\label{st}
\end{eqnarray}
stating, once again, the intensive proprieties of temperature.

Similarly, from definition (\ref{pres}) and assuming that the volumes scales, like an ordinary extensive quantity, in $V^{N{\cal M}}=N\,V^{\cal M}$, we obtain
\begin{eqnarray}
{1\over1+(q-1)\,\tilde\beta^{N{\cal M}}_{2-q}\,U^{N{\cal M}}}\left({\partial\,U^{N{\cal M}}\over\partial V^{N{\cal M}}}\right)_S={1\over1+(q-1)\,\tilde\beta^{\cal M}_{2-q}\,U^{\cal M}}\left({\partial\,U^{\cal M}\over\partial V^{\cal M}}\right)_S \ ,
\end{eqnarray}
that corresponds to the intensive propriety of the pressure
\begin{eqnarray}
\tilde P^{N{\cal M}}=\tilde P^{\cal M} \ .
\end{eqnarray}

We conclude by observing that, in addition to Eq. (\ref{energy}), other composition laws for the energy can be introduced consistently with the request of the factorization of distribution. For instance we can consider the following relation
\begin{eqnarray}
{\cal H}^{{\rm A}\cup{\rm B}}-U^{{\rm A}\cup{\rm B}}={\cal H}^{\rm A}-U^{\rm A}+{\cal H}^{\rm B}-U^{\rm B}+\omega\,\left({\cal H}^{\rm A}-U^{\rm A}\right)\,\left({\cal H}^{\rm B}-U^{\rm B}\right) \ ,\ \label{sume1}
\end{eqnarray}
explored in \cite{Ou}. It can be easily verified that this equation realizes the factorization of the distribution by using the second of Eq. (\ref{pdf}), when we pose $\omega=(q-1)\,\bar\beta_{2-q}$. On the other hand, Eq. (\ref{sume1}) can be rewritten equivalently in the form
\begin{eqnarray}
{\cal H}^{{\rm A}\cup{\rm B}}={\cal H}^{\rm A}+{\cal H}^{\rm B}-\omega\,{\cal H}^{\rm A}\,{\cal H}^{\rm B}+\omega\,\left({\cal H}^{\rm A}\,U^{\rm B}+{\cal H}^{\rm B}\,U^{\rm A}\right) \ , \label{sume2}
\end{eqnarray}
in accordance with Eq. (\ref{energy}), which, as already observed in \cite{Wang}, shows a dependence from the mean energy in the energy spectrum that is clearly unpleasant.

%%%%%%%%%%%%%%%%%%%%%%%%%%%%%%%%%%%%%%%%%%%%%%%%%%%%%%%%%%%%%%%%%
\section{Final discussion}

In this letter, we have shown that the definitions of temperature and pressure are sensitive to the composition law of entropy and internal energy governing the system. In the existent literature, usually the focus is mainly posed on the composition of entropy, neglecting the possible interactions between the parts of the system; a fact that has given rise to several questions about the consistency of the theory. An exception to this can be found in \cite{Wang1,Wang2} where was observed that the composition law (\ref{comp}) for the entropy necessarily induces a non-additive expression for the energy that is substantially equivalent to Eq. (\ref{sume}). In particular, Authors succeed to define, in a consistent way, several intensive variables in the framework of the incomplete probability distribution theory \cite{Wang3,Wang4}, leaving open the question about their definition in other formalisms.\\ In this paper, to overcome this lack, we introduce a definition of temperature and pressure, according to the zeroth principle of thermodynamics, when the system is governed by the entropy $S_{2-q}$ and the interactions are taken into account by means of Eq. (\ref{sume}). We remark one-more that the formalism used in this letter correspond to the first formalism of the Tsallis statistics operating with standard linear expectation values \cite{Tsallis1} although we have used a re-parameterized expression of the entropy which has the advantage to make the theory a little bit easy and symmetric in the various thermodynamics relations emerging in the theory.\\ Our main conclusions can be stated as follows: if the composition rule for the mean energy  (\ref{energy}) would be valid for real system, than one could study within non-extensive statistics the composition of systems and introduce in a Gibbsian manner intensive variables like temperature and pressure. Clearly, if this can be done, than the validity of the present approach is strongly limited to the validity of Eq. (\ref{energy}). Moreover, these definitions of intensive variables, whose intensive nature is consistent with the mathematical structure of the theory, are substantially different from other definitions presented in literature in the framework of the non-extensive statistical mechanics \cite{Abe,Abe1,Abe2,Wada,Wada3,Wada4,Scarfone0}.\\
Finally, since non-extensive thermodynamics is currently applied also to study problems outside the traditional physics (see, for instance, \cite{Takahashi} and references therein) we auspices that the present theoretical approach could better clarify the potential applications of this theory in other scientific disciplines.

%%%%%%%%%%%%%%%%%%%%%%%%%%%%%%%%%%%%%%%%%%%%%%%%%%%%%%%%%%%%%%%%%%%%%%%%
\appendix
\section{}
We show the invariance of the distribution (\ref{pdf}) for a uniform energy shift $\epsilon_i\to\epsilon_i'=\epsilon_i+\epsilon$. For sake of clarity, in the following we denote
\begin{eqnarray}
p_i^{(1)}={1\over\bar Z_{2-q}}\,\exp_q\Big(-\bar\beta_{2-q}\,(\epsilon_i-U)\Big) \ ,\label{pdf1}
\end{eqnarray}
and
\begin{eqnarray}
p_i^{(2)}={1\over\tilde Z_{2-q}}\,\exp_q\Big(-\tilde\beta_{2-q}\,\epsilon_i\Big) \ .\label{pdf2}
\end{eqnarray}
We recall, as discussed in the text, that all the quantities in the formulas (\ref{pdf1}) and (\ref{pdf2}) are related to Lagrange multipliers $\gamma$ and $\beta$, determined, in turn, by the corresponding constraint equations $\sum_ip_i=1$ and $\sum_ip_i\,\epsilon_i=U$, respectively. Let us observe that these equations are invariant under the uniform energy shift. In particular, for the energy constraint, we have
\begin{eqnarray}
\sum_{i=1}^Wp_i\,\epsilon'_i=U' \ ,\quad\Leftrightarrow\quad\sum_{i=1}^Wp_i\,(\epsilon_i+\epsilon)=U+\epsilon \ ,\quad\Leftrightarrow\quad\sum_{i=1}^Wp_i\,\epsilon_i=U \ .
\end{eqnarray}
As a consequence, if the distribution is invariant in form, the value of the Lagrange multipliers derived from the constraint equations do not change, i.e. $\beta=\beta'$ and $\gamma=\gamma'$.\\
Invariance in form of $p^{(1)}_i$ is trivial, by noting that $\epsilon'_i-U'=\epsilon_i-U$. This means that $\bar Z'_{2-q}=\bar Z_{2-q}$ and $\bar\beta'_{2-q}=\bar\beta_{2-q}$ so that $p_i^{(1)}$ is invariant, as known.

Differently, invariance in form of the distribution $p_i^{(2)}$ can be proved by using the relation
\begin{eqnarray}
\exp_q(x+y)=\exp_q(x)\,\exp_q\Big({y\over1+(1-q)\,x}\Big) \ ,
\end{eqnarray}
so that
\begin{eqnarray}
\nonumber
p_i^{(2)}&=&{1\over\tilde Z'_{2-q}}\,\exp_q\Big(-\tilde\beta'_{2-q}\,(\epsilon_i+\epsilon)\Big)\\
&=&{1\over\tilde Z''_{2-q}}\,\exp_q\Big(-\tilde\beta''_{2-q}\,\epsilon_i\Big) \ ,\label{pdf3}
\end{eqnarray}
where
\begin{eqnarray}
\tilde\beta''_{2-q}={\tilde\beta'_{2-q}\over1-(1-q)\,\tilde\beta'_{2-q}\,\epsilon} \ ,
\end{eqnarray}
and
\begin{eqnarray}
\tilde Z''_{2-q}=\tilde Z'_{2-q}\,\Big[1-(1-q)\,\tilde\beta'_{2-q}\,\epsilon\Big]^{1\over q-1} \ .
\end{eqnarray}
Equation (\ref{pdf3}) has the same form of $p_i^{(2)}$ given in Eq. (\ref{pdf2}). In this way, from the constraint equations, we obtain necessarily $\tilde Z''_{2-q}=\tilde Z_{2-q}$ and $\tilde\beta''_{2-q}=\tilde\beta_{2-q}$ stating the invariance of $p_i^{(2)}$ under energy shift translation.

%%%%%%%%%%%%%%%%%%%%%%%%%%%%%%%%%%%%%%%%%%%%%%%%%%%%%%%%%%%%%%%%%%%%%%%%%%

\vspace{10mm}

\noindent{\bf Acknowledgements}\\

\noindent The author is very grateful to Dr. E.K. Lenzi for helpful discussions
and critical reading of the manuscript.
%%%%%%%%%%%%%%%%%%%%%%%%%%%%%%%%%%%%%%%%%%%%%%%%%%%%%%%%%%%%%%%%%%%%%%%%

\vspace{10mm}
%\newpage

\noindent{\bf References}\\

\end{document}